# Comparing Example-Based Collaborative Reflection to Problem-Solving Practice for Learning during Team-Based Software Engineering Projects

Sreecharan Sankaranarayanan, Siddharth Reddy Kandimalla, Christopher Bogart, R. Charles Murray, Haokang An, Michael Hilton, Majd Sakr, Carolyn Rosé
sree@cmu.edu, skandima@andrew.cmu.edu, cbogart@andrew.cmu.edu, rcmurray@andrew.cmu.edu, haokanga@andrew.cmu.edu, mhilton@cmu.edu, msakr@andrew.cmu.edu, cprose@andrew.cmu.edu
Carnegie Mellon University, Pittsburgh, Pennsylvania, USA

**Abstract:** Contributing to the literature on aptitude-treatment interactions between worked examples and problem-solving, this paper addresses *differential learning from the two approaches when students are positioned as domain experts learning new concepts*. Our evaluation is situated in a team project that is part of an advanced software engineering course. In this course, students who possess foundational domain knowledge but are learning new concepts engage alternatively in programming followed by worked example-based reflection. They are either allowed to finish programming or are curtailed after a pre-specified time to participate in a longer worked example-based reflection. We find significant pre- to post-test learning gains in both conditions. Then, we not only find significantly more learning when students participated in longer worked example-based reflections but also a significant performance improvement on a problem-solving transfer task. These findings suggest that domain experts learning new concepts benefit more from worked example-based reflections than from problem-solving.

## Introduction

The trade-off between problem-solving practice and worked example study has not been deeply investigated in the software engineering context, especially for students who have moved beyond basic syntactic and semantic knowledge about programming and on to advanced topics such as Cloud Computing. In this context, the concepts and skills that students are learning are new, but they have acquired substantial foundational knowledge from their prior learning experiences. In domains where this trade-off between problem-solving practice and worked example study has been thoroughly investigated, extensive problem-solving practice is generally considered inferior for positively impacting student learning (Renkl, 2014). Contrary to what this might suggest for software engineering also, problem-solving practice (i.e., computer programming) has remained the predominant form of pedagogy. This may be because the literature on computer science education does not provide a definitive answer about this trade-off or that the findings are thought to apply especially to conceptual knowledge, and not to performance on more authentic, complex problem-solving tasks. Studies adjacent to the worked example literature relying still on cognitive load theory have variously found positive effects (Margulieux et al., 2012) as well as mixed effects (Morrison et al., 2015) in the software engineering context warranting further study about this important comparison. We seek, therefore, to contribute to the literature on aptitude-treatment interactions between worked examples and problem-solving by addressing the fundamental question of *how students learn differentially from the two when they are positioned as domain experts learning new concepts and skills*. We position our investigation in the software engineering context. We are also especially interested in the follow-up question of whether they might gain more conceptual knowledge from worked example study but be left less able to engage effectively in subsequent problem-solving.

To answer these questions, we conduct our study in a synchronous, online, team-based software engineering course on Cloud Computing for graduate students and advanced undergraduate students. We assign students, in their teams, to two conditions. In the first, which we call the *maximize learning from problem-solving (MLPS) condition*, teams are tasked to complete problem-solving and then engage in a brief collaborative reflection based on a worked example in the remaining time. In the second, called the *maximize learning from reflection (MLR) condition*, teams are curtailed from problem-solving after a pre-specified amount of time regardless of whether they reach a completed solution. They subsequently engage in a full-length reflection based on the worked example. The difference is where the time boundary is placed between problem-solving and collaborative reflection. The results challenge deeply held assumptions in computer science education about the extensive computer programming practice being an activity necessary for student learning.

## Method

### Course Context

This study was conducted in a graduate-level project-based online software engineering course on Cloud Computing offered to graduate and advanced undergraduate students at Carnegie Mellon University and its branch campuses. The course is structured around five project-based units. Each unit has several sub-units and culminates in a large individual project that has assessment components to evaluate achievement in each sub-unit. Our experiment is situated within unit 3.3 that focuses on *"multi-threaded programming and consistency"*. In this sub-unit, students, in groups of 4, work with our synchronous collaborative programming activity, called the Online Programming Exercise (OPE). A summary of the course structure and the location of the study within it is shown in Figure 1. A total of 74 students completed the exercise and the subsequent project. Enrollment numbers were about half the usual as a result of the COVID-19 pandemic. No other substantial changes in course content or structure were needed since the course had been offered online for over 10 prior semesters.

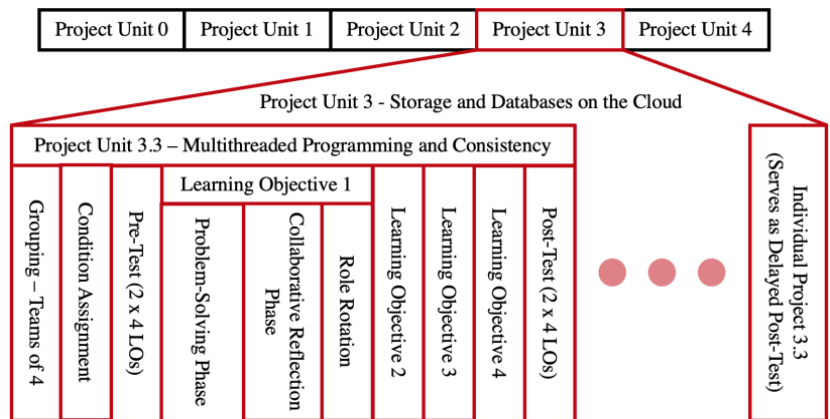

Figure 1: Course structure, pre-test, post-test, and delayed post-test alignment.

### Design of the Online Programming Exercise (OPE)

The collaborative programming exercise is divided into four tasks, each targeting a learning objective (LO). Each task is divided into a *problem-solving phase* and a *collaborative reflection phase.* During the problem-solving phase, students are assigned to four independent roles (Driver, Navigator, Researcher, Project Manager) based on an instructional adaptation of the industry practice of Mob Programming (Sankaranarayanan, 2019; Sankaranarayanan et al., 2019). In the subsequent collaborative reflection phase, they are guided by conversational agent-based prompts to reflect based on a presented worked example. The prompting infrastructure is based on the open-source Bazaar conversational agent framework (Adamson et al., 2014). The roles that students are assigned to rotates after each task. In the pre- and post-tests that students complete immediately before and after the task, respectively, they attempt two multiple-choice questions per LO. Performance improvement from pre- to post-test averaged per LO is used as a measure of students' conceptual learning from the task. The individual programming project that serves as a procedural and conceptual delayed post-test is graded by the instructor on a rubric with 12 quality scores, each of which ranges between 0 and 1. Table 1 shows the learning objectives, examples of pre- and post-test questions, and conversational agent-based collaborative reflection prompts corresponding to each task, while Figure 1 shows the position of pre-, post-, and delayed post-tests within the course.

Table 1: Learning Objectives, Corresponding Pre/Post Test Questions (Examples), Information and Transactivity Prompts

|   | Learning Objective | Example Pre/Post Test Question - Multiple-Choice | Example Collaborative Reflection Prompts |
|---|---|---|---|
| 1 | Building blocks of multithreading. | Which of the following statements about multithreading in Java is INCORRECT? | Was your approach similar to the reference solution? What Thread class functions did you use? Take turns explaining the logic. |
| 2 | Diagnosing and fixing deadlocks. | The usage of notify() will never result in a deadlock in which of the following multithreaded scenarios? | In an ideal scenario, can you think of a built-in Java thread-safe class that could replace the priority queue? Take turns explaining. |

| 3 | Diagnosing and preventing a race condition. | Examining the following code snippets, identify the one that will NEVER lead to a race condition. | Comparing your approach to the reference solution, how did you avoid the race condition here? Take turns explaining the logic. |
|---|---|---|---|
| 4 | Ensuring strong consistency in data stores. | How would you acquire a lock on a critical resource shared by multiple threads to ensure consistent runtime behavior? | Can you put what you are learning in all these tasks together to think about ensuring strong consistency? Take turns explaining. |

### Experimental Design

Two weeks before the experimental manipulation, students participated in a training OPE session in randomly formed teams of 4 based on their time availability. In preparation, students were provided with videos and text explaining the OPE and motivating its use for collaborative team projects. The exercise was relatively simple data processing using the 'pandas' library in Python. While still a meaningful component of the course, it was meant as an opportunity for students to familiarize themselves with role-taking, role-rotation, collaborative reflection, and the interface of the Cloud9 IDE used for the task. Each exercise session lasted for a total of 80 minutes.

For the experimental manipulation, students were again grouped randomly into teams of 4 based on their time availability while ensuring that they weren't placed into teams with students they had done the training with. The activity, once again, lasted a total of 80 minutes. A total of 74 students were assigned to 19 teams of which 17 were 4-member teams and 2 were 3-member teams. In the 3 member teams, the student assigned to the project manager role also acted as the researcher. 9 teams were assigned to the *maximize learning from problem-solving (MLPS) condition*, where for each task, teams complete the problem-solving and then enter into a reflection phase for the remaining time, and 10 teams were assigned to the *maximize learning from reflection (MLR)* condition, where problem-solving was curtailed after a pre-specified amount of time, and they enter the reflection regardless of whether they completed the problem-solving or not.

## Hypotheses, Analysis, and Results

*Hypothesis 1: The Online Programming Exercises (OPEs) results in pre- to post-test learning gains*

To evaluate the general value for learning of the activity regardless of condition, we compared pre- and post-test scores per learning objective, role, and condition, where pre- and post-test scores vary between 0 and 1 per learning objective. For this analysis, we build an ANOVA model with test score as the dependent variable, and time-point (pre- vs post-test), condition (MLPS vs MLR), role (Driver, Navigator, Researcher, Project Manager), and learning objective (listed in Table 1) as independent variables. We also included pairwise interaction terms between time-point and each of the other three independent variables. There was a significant effect of time-point $F(1,410) = 3.77$, $p < .0001$, effect size .38 s.d., with an average pre-test score of .55 (.37 s.d.) and average post-test score of .69 (.37 s.d.). None of the pairwise interactions were significant. Thus, we confirmed that students in both conditions learned based on the significant difference between pre- and post-test scores across the two conditions regardless of role or learning objective. Thus, the first hypothesis is confirmed.

Hypothesis 2: *The MLR condition will result in better pre- to post-test learning gains.*

In order to test the effect of condition on the magnitude of learning we compared post-test scores between conditions controlling for pre-test scores. In particular, we computed an ANCOVA model with post-test score as dependent variable, pre-test score as the covariate, and condition (MLPS vs MLR), and learning objective as independent variables. We found a significant effect of condition such that students in the MLR condition learned more $F(1,254) = 6.0$, $p < .05$, effect size .24 s.d.. For the MLR condition, the average pre-test score was .53 (.36 s.d.) and post-test score was .72 (.34 s.d.), and for the MLPS condition, average pre-test score was .57 (.38 s.d.) and post-test score was .65 (.37 s.d.) We find significantly higher pre- to post-test gains in the MLR condition in comparison with the MLPS condition ($p < 0.05$), which suggests that students with domain expertise benefit more from the worked example-based reflection than the problem-solving for acquisition of new conceptual knowledge. Thus, the second hypothesis is confirmed.

Hypothesis 3: *The MLPS condition will result in better performance on the delayed post-test*

In order to test the effect of condition on achievement on the transfer task, we considered each of the 12 quality ratings assigned by the instructor within a single model in order to control for multiple comparisons. In particular, we computed a single ANCOVA model with numeric quality rating as the dependent variable, total pre-test score across learning objectives as a covariate, and the condition and the name of the quality rating as independent variables. We also include the pairwise interaction term between the two independent variables. There was a significant effect of condition such that students in the MLR condition scored higher than students in the MLPS condition, $F(1,707) = 4.36$, $p< .05$, effect size .15 s.d., which is a weak effect. The average score was 4.7 (2.7 s.d.) for the MLPS condition and 5.0 (2.5 s.d.) for the MLR condition. There was no significant interaction between condition and quality rating name. Thus, although the effect is weak, students in the MLR condition performed better than students in the MLPS condition across the 12 quality ratings. The third hypothesis is rejected, and in fact, the opposite is supported.

## Discussion and Conclusion

In this paper we presented a study in which we contrasted two conditions: *maximize learning from problem-solving* and *maximize learning from reflection*.

First, we find that the team project exercises lead to significant pre- to post-test learning in both conditions. This indicates that both worked example study, and problem-solving practice are potentially valuable for learning. While we did not compare the sequential presentation of the problem-solving, worked examples and the collaboration scaffolds with either of the scaffolds provided on their own, the lack of a detrimental effect in either condition means that the role of the scaffolds was not so redundant as to draw student attention away from the relevant problem states.

When comparing across conditions, we see that the condition where students spent more time on worked example-based reflection resulted in significantly more pre- to post-test learning. Based on cognitive load theory, we could surmise that it is indeed the case that extensive problem-solving consists of production steps that are superfluous to the learning here. While problem-solving practice was not detrimental to student learning, we can more efficiently use student time and impact their learning more if we use worked examples as well, with an emphasis on time spent on reflecting rather than the completion of the problem-solving.

One concern among educators has been that while students' conceptual learning can be positively impacted by the use of worked examples, they may not perform as well when asked to problem-solve on a transfer task because they received less practice. We started with the hypothesis that this might be the case, and we would have not been surprised to have found that. However, what we found was the opposite. Students who reflected longer also performed better on a subsequent authentic programming task, though the effect size was small. We can conclude that students, at this point in the course, had already acquired the procedural knowledge of programming enough to not need the practice i.e., given a schema, they were able to translate that into a solution to the problem. The positive impact of the worked example condition on the conceptual process of schema acquisition and induction then led to a positive impact on student performance on the subsequent project also.

## Acknowledgements

This work was funded in part by NSF grants IIS 1822831, IIS 1917955 and funding from Microsoft.